\documentclass[twocolumn,showpacs,preprintnumbers,amsmath,amssymb]{revtex4}
\usepackage{graphicx}
\usepackage{epsfig}
\usepackage{dcolumn}
\usepackage{amssymb}
\usepackage{color}
\usepackage{hyperref}
\usepackage{bm}
\usepackage{amsfonts}
\usepackage{ulem}

\def\be{\begin{equation}}
\def\ee{\end{equation}}
\def\bq{\begin{eqnarray}}
\def\eq{\end{eqnarray}}

\def\bm{\begin{multicols}{2}}
\def\em{\end{multicols}}
\def\ed{\end{document}}
 1

\begin{document}

\title{Quantum phases of 1D Hubbard models with three- and four-body couplings}

\author{Fabrizio Dolcini${}^{1,2}$}

\author{Arianna Montorsi${}^1$}
\email[]{arianna.montorsi@polito.it}
\affiliation{${}^1$ Dipartimento di Scienza Applicata e Tecnologia del Politecnico di Torino\\ corso Duca degli Abruzzi
24, 10129 Torino, Italy \\
${}^2$ CNR-SPIN, I-80126 Napoli, Italy}
\date{\today}

\begin{abstract}
The experimental advances in cold atomic and molecular gases stimulate the investigation of lattice correlated systems beyond the conventional on-site Hubbard approximation, by possibly including multi-particle processes. We study fermionic  extended Hubbard models in a one dimensional lattice with different types of particle couplings, including also three- and four-body interaction  up to nearest neighboring sites. By using the Bosonization technique, we investigate the low-energy regime and determine the conditions for the appearance of ordered phases, for arbitrary  particle filling. We find that three- and four-body couplings may significantly modify the phase diagram. In particular, diagonal three-body terms that directly couple the local particle densities  have qualitatively different effects from off-diagonal  three-body couplings originating from correlated hopping, and favor the appearance of a Luther-Emery phase even when  two-body terms are repulsive. Furthermore, the four-body coupling gives rise to a rich phase diagram and may lead to the realization of the Haldane insulator phase at half-filling.

\end{abstract}
\maketitle

\section{\bf Introduction} 
The   recent experimental realization of ultracold fermionic gases of molecules \cite{NI} and atoms \cite{LBL} with strong dipolar moments, and their confinement in optical lattices \cite{CHO}  allow  to  investigate in a controlled way  the effect of interaction in one-dimensional (1D) lattice  systems. In such dimension correlations are well known to be relevant, so that 1D systems are characterized by peculiar properties that cannot be captured by the ordinary Fermi liquid theory. As compared to the traditional solid-state realizations of 1D systems  such as Bechgaard salts~\cite{BECH}, 1D cuprates~\cite{CUPR}, semiconductor quantum wires\cite{QW}, carbon nanotubes~\cite{CNT}, and edge states in Quantum Hall effect systems~\cite{QHE}, optical lattice based implementations allow for a greater tunability of the interaction parameters  and for the implementation of peculiar types of interactions, such as long-range  or many-body couplings~\cite{LEW-review,LLCA,LEW}. These features represent a remarkable boost in the investigation of correlations in 1D systems, broadening the range of accessible parameters, and the spectrum of physical properties that can be addressed.

The prototype Hamiltonian utilized to account for correlation effects in lattice fermion systems is  the Hubbard model, originally introduced in the context of condensed matter physics~\cite{HUB}. It describes electron-electron interaction as a purely {\it on-site} repulsion between electrons with opposite spin orientation. Despite its simplicity, it does show that Coulomb interaction has dramatic effects on the electron dynamics in low dimensions, leading the 1D   electronic system   into an insulating state (the Mott insulator, MI) at half-filling, no matter how weak the repulsion is. Such phase was realized with controlled systems of neutral ultracold fermionic atoms \cite{MOTT}, where an arbitrary on-site interaction is obtained via appropriate Feshbach resonance.
Very recently, in these systems  it has also become possible to simulate  longer range  couplings, thanks to the confinement of systems of molecules with non vanishing dipolar moment. 
This leads to consider generalizations of the Hubbard model, including further interactions terms characterized by various coupling constants, such as nearest neighbors density-density coupling, correlated hopping, exchange interaction and so on. Such models, often referred to as the class of  extended Hubbard Hamiltonians, have been adopted in the description of  various phenomena in condensed matter~\cite{HIRSCH,AAch,PK,JAKA3,CC,SAG}.\\

A quite promising research frontier for the investigation of Hubbard Hamiltonians is opened by the study of ultra cold atoms and molecules. 
Indeed the tunability of the various coupling constants  in such atomic and molecular systems is  easier than in condensed matter physics.
Moreover, these systems have spurred the interest in the role of three-body interaction terms, which are often disregarded in condensed matter problems. 
It has for instance been predicted that polar molecules in optical lattices of various geometries naturally give rise to Hubbard models with strong nearest neighbour three-body interactions, which can be controlled in a independent way from  the two-body terms~\cite{BMZ, Bonnes, hammer}. 
An experimental evidence of the role of three-body interaction has been observed in cold ${}^{85} {\rm Rb}$ Rydberg atoms trapped in a magneto-optical trap\cite{han}. Furthermore, recent observations on cold {\rm Cs} atoms have provided the signature that even four-body interactions affect the level population~\cite{gurian}.
These experimental advances pave the way to the search for other phases than the Mott insulator. Indeed, it is known that three- and many-body terms are strong candidates for the observation of exotic phases, such as fractional quantum Hall states in electron systems~\cite{MORE}. More recently, it has been found that, in the case of bosonic particles,  solid and supersolid phases are favored \cite{BMZ,CSal} in the strong three-body regime.  In the fermionic case, a three-body correlated hopping was predicted \cite{BMR} to favor Haldane charge order in principle at half-filling. Such terms are off-diagonal in the occupation number representation though. \\

An exhaustive characterization of the phase diagram of the extended Hubbard models, in particular in the presence of diagonal three- and four-body interaction terms, is thus lacking. This article is devoted to the investigation of this problem. 
In Sec.\ref{sec-2} we consider a quite general class of fermionic Hubbard models, which includes various types of nearest neighbors two-body, as well as three- and four-body interaction terms,   characterized by independent coupling constants. We make use of the  Bosonization Technique to investigate the low-energy limit, and we  determine the conditions on the coupling constants for the onset of different phases (sec.\ref{sec-3}), such as Luttinger liquid, the Luther-Emery liquid, charge insulators (Mott and Haldane) and fully gapped phases. Then, in Sec.\ref{sec-4}, we focus on the effect of three-body and four-body terms. We show that diagonal three-body couplings have quite different effects on the phase diagram from the off-diagonal three-body coupling that were investigated in the correlated hopping models. In particular they favor the presence of the Luther-Emery phase, even in the presence of repulsive on-site interaction. Then, we show that in the presence of four-body interaction the phase diagram acquires an extremely rich structure, where a large variety of phases can be obtained with varying the coupling constants, including the Haldane insulator phase. Finally, in sec.\ref{sec-5} we summarize and discuss our results.

\section{Model Hamiltonian and its low energy limit}
\label{sec-2}
The  Extended Hubbard model  that we consider is described by the following Hamiltonian 
\begin{widetext}
\begin{eqnarray}
{\cal{H}}& =&   \sum_{j=1}^{N_s}  \left[ -\sum_{\sigma} (c_{{j},\sigma}^\dagger
c_{j+1, \sigma} + h.c.)[t-X (\hat{n}_{j
\bar{\sigma}}
+ \hat{n}_{j+1 \bar{\sigma}})+\tilde{X} \, \hat{n}_{j \bar{\sigma}}
\hat{n}_{j+1 \bar{\sigma}} ]  \, \, \,+ U \, \hat{n}_{j \uparrow}\hat{n}_{j \downarrow} + V \, \hat{n}_{j} \hat{n}_{j+1} + \right.\label{EHM} \\ & & \left. \hspace{1cm} +J \,
{\mathbf{S}}_{j} \cdot {\mathbf{S}}_{j+1} + Y (c_{j\uparrow}^\dagger
c_{j \downarrow}^\dagger  c^{}_{j+1 \downarrow} c^{}_{j+1 \uparrow}
+ {\rm h.c.} )+ P \, (\hat{n}_{j \uparrow} \hat{n}_{j \downarrow} \hat{n}_{j+1}+
\hat{n}_{j+1 \uparrow} \hat{n}_{j+1 \downarrow} \hat{n}_{j})
+ Q \,  \hat{n}_{j \uparrow} \hat{n}_{j \downarrow} \hat{n}_{j+1 \uparrow}
\hat{n}_{j+1 \downarrow} \right] \quad.
\,  \nonumber
\end{eqnarray}
\end{widetext}
In Eq.(\ref{EHM}) $N_s$ denotes the number of sites of the 1D lattice, $c^{\dagger}_{j \sigma}$ and $c^{}_{j \sigma}$ are fermionic creation and annihilation operators, $\sigma=\uparrow, \downarrow$ being the spin label; $\hat{n}_{j \sigma}=c^{\dagger}_{j \sigma} c^{}_{j \sigma}$ is the fermion number operator for spin~$\sigma$ at the $j$-th lattice site, and $\hat{n}_j=\hat{n}_{j \uparrow}+\hat{n}_{j \downarrow}$; finally ${\mathbf{S}}_{j}=\sum_{\sigma \sigma^\prime}(c^{\dagger}_{j \sigma} \, \mathbf{\boldsymbol\sigma}_{\sigma \sigma'} c^{}_{j \sigma'})/2$ is the spin operator ($\mathbf{\boldsymbol\sigma}$ are the Pauli matrices).
In Eq.(\ref{EHM}) $t>0$ represents the hopping amplitude for electrons, while $U$ is the customary Hubbard on-site interaction term. The couplings $X$ and $\tilde{X}$ account for correlated hopping terms, which have been first considered by Hirsch\cite{HIRSCH} and later by Simon and Aligia~\cite{AAch} in modeling hole superconductivity in narrow-band materials. Furthermore $V$ is the neighboring site density-density interaction, $J$ characterizes the exchange coupling, important in describing the onset of magnetic phases\cite{CUPR}, and $Y$ is a pair-hopping term which was first introduced by Penson and Kolb to provide an effective description of short-radius pair superconductivity (see Refs.\cite{PK,JAKA3}). Then,  $P$ parametrizes a three-body interaction   that directly couples the local fermonic densities. Thus, differently from the three-body term $\tilde{X}$ appearing in the correlated hopping, the $P$ term is diagonal  in the occupation number representation. Finally,   in Eq.(\ref{EHM}) the $Q$ term  describes a four-body interaction. In condensed matter systems such many-body terms,  although not stemming directly from Coulomb interaction,   can appear  indirectly as effective terms on decimated lattices, for instance in the study of Metal-Insulator transitions\cite{CC}, as well as in the mapping of a three-band Hubbard model into an effective single band\cite{SAG,NOTE}. In polar molecule systems, instead,  such terms are more directly realizable and can in principle be tuned independently from the two-body interaction. \\
The Hamiltonian (\ref{EHM}) can be easily verified to exhibit the total spin $SU(2)$  as well as the charge $U(1)$ symmetries, whereas further symmetries appear for specific relations between the coupling constants~\cite{DOMO} \\

\noindent In the weak coupling regime (i.e. for $|U,V,X,\tilde{X},J,Y,P,Q| \ll 4t$) one can fairly capture the physics of the lattice model by linearizing the dispersion relation of the hopping term near the two symmetric Fermi points $\pm k_F$, and by passing to the continuum limit through the replacement
\begin{equation}
c^{\dagger}_{j \sigma} \, \rightarrow \, \sqrt{a} \, \left(e^{-i k_F x} R^{\dagger}_{\sigma}(x)+e^{+i k_F x} L^{\dagger}_{\sigma}(x)\right) \label{climit} \quad,
\end{equation}
where $a$ is the original lattice spacing and $x=j a$. The fields $R_\sigma(x)$ and $L_\sigma(x)$ respectively describe the right and left moving components of the fermions, and are supposed to be slowly varying over distances of the order of $a$.
According to abelian bosonization~\cite{BOOK,EMERY,HALD,SHANKAR,STONE,VOIT,SOL,DELFT,GIAMARCHI}, these fermionic fields can be rewritten in the following way:
\begin{eqnarray}
R^{\dagger}_{\sigma}(x) &=& \frac{\kappa_{R \sigma}}{\sqrt{2 \pi \alpha}}
\exp{[-i \sqrt{4 \pi} \, \frac{\Phi_{\sigma}(x)+\Theta_{\sigma}(x)}{2}] } \label{R}  \\
L^{\dagger}_{\sigma}(x)&=&\frac{\kappa_{L \sigma}}{\sqrt{2 \pi \alpha}}
\exp{[+i \sqrt{4 \pi} \, \frac{\Phi_{\sigma}(x)-\Theta_{\sigma}(x)}{2}] }   \label{L}
\end{eqnarray}
where $\Phi_{\sigma}(x)$ and $\Theta_{\sigma}(x)$ are bosonic fields, mutually non-local and fulfilling $[ \Phi_{\sigma}(x), \Theta_{\sigma'}(y)]=\delta_{\sigma, \sigma'} {\rm sgn}(x-y)/2$. In Eqs.(\ref{R})-(\ref{L}) $\kappa_{R \sigma}, \kappa_{L \sigma}$ are Majorana Klein factors accounting for anticommutation of different fermionic species; $\alpha$ is an ultraviolet cut-off, which is of the order of the lattice spacing~$a$. 
Applying the Bosonization scheme  (see App.\ref{App}), and 
introducing the charge(c) and spin(s) fields $\Phi_{c/s}(x)=(\Phi_{\uparrow}\pm \Phi_{\downarrow})/\sqrt{2}$, the Hamiltonian (\ref{EHM}) exhibits in the low-energy limit the  charge-spin separation, namely it can be rewritten as the sum
\begin{equation}
{\cal{H}}={\cal{H}}_{c}+{\cal{H}}_s \quad, \label{Hcs}
\end{equation}
where ${\cal{H}}_\nu$ ($\nu=c,s$) is a Sine-Gordon model
\begin{eqnarray}
{\cal{H}}_\nu &=& \hbar v_\nu \int dx   \, [  \frac{:{\Pi_{\nu}^{\prime}}^2:+: (\partial_x
{\Phi_{\nu}^{\prime}})^2 :}{2}+ \nonumber \\
& & \hspace{2cm} +m_{\nu}\frac{
\cos{(\sqrt{8 \pi K_\nu} \Phi_{\nu}^{\prime}(x))}}{2 \pi \alpha^2} ] \quad. \label{HBOS}
\end{eqnarray}
Here $v_{c(s)}$ is the velocity of charge(spin) excitation along the chain, $\Pi_{\nu}^{\prime}(x)=\partial_x \Theta_{\nu}^{\prime}(x)$ is the  momentum conjugate to $\Phi_{\nu}^{\prime}$, where $\Phi_{\nu}^{\prime}=\Phi_{\nu}/\sqrt{K_{\nu}}$ and $\Theta_{\nu}^{\prime}=\Theta_{\nu} \sqrt{K_{\nu}}$ are the fields renormalized by the interaction; finally $m_{\nu}$ is the (dimensionless) mass parameter. 
\\The parameters  $K_\nu$, $v_\nu $ and $m_\nu$ appearing in Eq.(\ref{HBOS}) are determined by the following relations
\begin{equation}
\left\{ \begin{array}{lcl}
\displaystyle  v_\nu K_\nu &=& \displaystyle  v_F^0 (1+A_\nu) \\ & & \\
\displaystyle \frac{v_\nu}{K_\nu} &=& \displaystyle  v_F^0 (1+A_\nu-B_\nu) \\ & & \\
\displaystyle  v_\nu m_\nu &=& \displaystyle  v_F^0 \, C_\nu \end{array} \right. \label{relations}
\end{equation}
\\where $v_{F}^{0}= 2t a \hbar^{-1} \sin(\pi \rho/2)$ is the Fermi velocity in the non-interacting case, $\rho=2k_Fa/\pi$ is the filling factor, and the dimensionless quantities $A_\nu$, $B_\nu$ and $C_\nu$ read
\begin{equation}
A_{c}=[-X \rho+\tilde{X} \, (\frac{\rho^2}{4}-\frac{\sin^2\frac{\pi \rho}{2}}{\pi^2})-\frac{2Y}{\pi} \sin{\frac{\pi \rho}{2}}]/t \label{Ac} \\
\end{equation}
\begin{equation}
A_{s}=[-X \rho+\tilde{X} \, (\frac{\rho^2}{4}-\frac{\sin^2\frac{\pi \rho}{2}}{\pi^2})+\frac{J}{\pi} \sin{\frac{\pi \rho}{2}}]/t , \label{As}
\end{equation}
\begin{widetext}
\begin{eqnarray}
B_c &=&-\frac{1}{2\pi t \sin(\frac{\pi \rho}{2})} \left[U +2 V (2-\cos{\pi \rho})+ 8 X \cos{\frac{\pi \rho}{2}}-4 \tilde{X} \left( \, \rho \cos{\frac{\pi \rho}{2}}+\frac{\sin(\frac{\pi \rho}{2}) -\sin(\frac{3 \pi \rho}{2})}{\pi} \right)+2 Y-\frac{3}{2} J \cos{\pi \rho}+ \right. \nonumber \\
& & \left. \hspace{2cm} +  2 P \left(\rho (3-\cos\pi \rho)-\frac{2}{\pi} \sin{\pi \rho}\right)+
Q \left(\frac{\rho^2}{2} (3-\cos{\pi \rho})-\frac{2 \rho}{\pi} \sin{\pi \rho}-
\frac{1-2 \cos{\pi \rho}+\cos{\frac{\pi \rho}{2}}}{\pi^2} \right) \right] \label{Bc} \\
 & & \nonumber \\ 
 & & \nonumber \\
B_s &=& \frac{1}{2\pi t \sin(\frac{\pi \rho}{2})} \left[ U +2 V \cos{\pi \rho}+ 8 X \cos{\frac{\pi \rho}{2}}-4 \tilde{X} \left( \rho \cos{\frac{\pi \rho}{2}}-\frac{2}{\pi} \sin{\frac{\pi \rho}{2}} \right)+2 Y-\frac{J}{2}(2+ \cos{\pi \rho}) \, + \right. \nonumber \\
& & \left. \hspace{2cm}+ 2 P \left(\rho (1+\cos{\pi \rho})-\frac{2}{\pi} \sin{\pi \rho}\right)+
Q \left(\frac{\rho^2}{2} (1+\cos{\pi \rho})-\frac{2}{\pi} \rho \sin{\pi \rho}+
\frac{2}{\pi^2} (1-\cos{\pi \rho})) \right) \right] \label{Bs}
\end{eqnarray}
\end{widetext}
and
\begin{eqnarray}
C_c &=& \displaystyle -\frac{U-2 V-\frac{8 \tilde{X}}{\pi}-  2 Y +\frac{3}{2}J- \frac{4 Q}{\pi^2}}{2\pi t } \, \, \delta_{\rho,1} \label{Cc} \\
C_s&=&B_s \quad .\label{Cs}
\end{eqnarray}
The relation  $B_s=C_s$ in Eq.(\ref{Cs}) directly stems from the spin-$SU(2)$ symmetry of the model. Notice that $B_s$   depends on the value of filling $\rho$. In contrast, in the charge sector, $B_c \neq C_c$ in general. Furthermore, while $B_c$ is present at arbitrary filling, the term $C_c$ in Eq.(\ref{Cc}) is present only at half-filling ($\rho=1$), and originates from two-particle Umklapp processes. We have neglected here the presence of higher order terms in the Bosonization expression that may,   at specific commensurate filling values different from $\rho=1$,  couple charge and spin sectors\cite{nota-commensurate}. These effects have been investigated for instance in Refs.\cite{kolomeisky, nakamura,BMR}.\\

\noindent In the weak coupling regime, one has $|A_\nu|,|B_\nu|, |C_\nu|  \ll 1$ and the general relations (\ref{relations}) imply
\begin{eqnarray}
K_\nu&\simeq  &   \, 1+\frac{B_\nu}{2} \label{Knu} \\
m_\nu&\simeq  &    \,C_\nu \label{Cnu} \quad \quad.
\end{eqnarray}
\section{Quantum phase diagram}
\label{sec-3}
An important criterion to classify the various phases is to identify the presence of charge and/or spin gaps. Due to the spin-charge separation  (\ref{Hcs}), this task is accomplished by analyzing the Sine-Gordon model (\ref{HBOS}) charactering each of the two sector.   The Sine-Gordon model, for which the exact solution is  known\cite{SGM}, may give rise to  gapless and gapped phases, depending on the range of the parameters $K_\nu$ and $m_\nu$.  Its asymptotic properties can be captured through RG analysis, which yields the following RG flow equation
\begin{equation}
\frac{d \xi_{\nu}}{d l}=-\eta_{\nu}^2 \hspace{1cm} \frac{d \eta_\nu}{d l}=-\xi_{\nu} \eta_{\nu} \label{RGeq}
\end{equation}
where $\xi_\nu=4(\sqrt{K_\nu}-1) \simeq B_\nu$ and $\eta_\nu=m_\nu\simeq C_\nu$ are dimensionless space parameter coordinates. The RG flow, characterized by the scaling invariant $\xi_\nu^2-\eta_\nu^2 = \mbox{const}$, shows that the model is gapless if and only if the bare parameters belong to the region $\xi_\nu \ge |\eta_\nu|$~\cite{nota-RG}.
In particular, for the spin sector ($\nu=s$), the spin-SU(2) symmetry of the model, Eq.(\ref{Cs}), causes the RG flux   to take place along the separatrices $\xi_s=\pm \eta_s$ of Eq.(\ref{RGeq}), so that $K_s$ and $m_s$ cannot  vary independently.\\

In the gapless phase, the fields $\Phi_\nu (x)$ oscillate and --except for the case of Umklapp processes-- the integral of the cosine term in (\ref{HBOS})  vanishes on average. In contrast, the opening of a gap takes place whenever the vacuum expectation value $\langle \Phi_\nu \rangle$ of the corresponding field pins to a value that minimizes the cosine in Eq.(\ref{Hcs})~\cite{GIAMARCHI,JAKA1}. As far as the   charge sector is concerned, a   gap can open only at half-filling, and there are two possible sets of pinning  values for $\Phi_c$, depending on whether $m_c<0$ or $m_c>0$. If the spin gap is closed, these values correspond to the two possible phases of a charge insulator, which are denoted as Mott insulator (MI) and Haldane insulator (HI), respectively, for reasons that will be clarified below. In contrast, for the spin sector the SU(2) invariance makes the opening of a spin gap always correspond to $m_s<0$, so that only one way of pinning $\Phi_s$ is possible. In this case, when the charge gap is closed the spin gapped phase is the Luther-Emery (LE)  phase, whereas when  the charge gap is also open one has two possible fully gapped phases: for $m_c<0$  the bond ordered wave phase (BOW) and for $m_c>0$ the charge density wave (CDW) phase.
The transitions from the gapless phase to the gapped phases are of Berezinsky-Kosterlitz-Thouless (BKT) type, whereas the transition from the $m_c>0$ to the $m_c<0$ phase is of second order\cite{SAC}. The possible scenarios are summarized in   table (\ref{table1}), where the behavior of the fields $\Phi_c(x)$ and $\Phi_s(x)$ is described. Depending on closing/opening of the related gap $\Delta_\nu$ ($\nu=c,s$), the field $\Phi_\nu$ can either be fluctuating or pinned around a set of values indicated in the table and characterized by integers $p_\nu$.\\

\be
\begin{tabular}{|c|c|c|c|c|clcl}
\hline
$\Delta_c$ & $\Delta_s$  & $  \sqrt{2\pi} \Phi_c $ & $ \sqrt{2\pi}\Phi_s $ & Type of Phase  \\
\hline \hline
=0 & =0 & \mbox{\small fluctuating} & \mbox{\small fluctuating} & LL \\ \hline
=0 & $\neq 0$ & \mbox{\small fluctuating} & \mbox{\small  $\pi p_s$} & LE \\ \hline
$\neq 0$ & $=0$  & \mbox{\small $\pi p_c $} & \mbox{\small fluctuating} & MI \\ \hline
$\neq 0$ & $=0$  &  \mbox{\small $\pi (p_c+1/2)$  } & \mbox{\small fluctuating} & HI \\ \hline
$\neq 0$ & $\neq0$ & \mbox{\small $\pi p_c$} & \mbox{\small $\pi p_s $} & BOW \\ \hline
$\neq 0$ & $\neq0$ & \mbox{\small $\pi (p_c+1/2)$  } & \mbox{\small $\pi p_s$} & CDW \\
\hline
\end{tabular}\,   \,\label{table1}
\ee
Each of the above gapped phase is characterized by a specific LRO \cite{MORO,BMR}. To see that, one can introduce --already at the level of the lattice model-- the parity and string operators at a given site $j$, i.e. {\it non-local} operators defined as

\begin{eqnarray}
O_P^{(\nu)}(j) &=& \displaystyle \prod_{l=1}^{j} e^{i \pi n J_{l}^{(\nu)}} \quad   \\
\quad O_S^{(\nu)}(j) &=& \displaystyle   \prod_{l=1}^{j} e^{i\pi n J_{l}^{(\nu)}} J_{l}^{(\nu)} ,
\end{eqnarray}
respectively, with $\nu=c,s$, and $J_{l}^{( c)}=(\hat{n}_{l}-1)$,  $J_{l}^{(s)}=(\hat{n}_{l,\uparrow}-\hat{n}_{l,\downarrow})$. The two-point correlators $C_{P}^{(\nu)}( r)\doteq \langle O_P^{(\nu)} (j) O_P^{(\nu)\dagger} (j+r) \rangle$
(parity correlator), and $C_{S}^{(\nu)}( r) \doteq \langle O_S^{(\nu)} (j) O_S^{(\nu) \dagger} (j+r)\rangle $
(string correlator)    can be evaluated in the continuum limit  along the lines of Refs. \cite{GIAMARCHI,MORO}, obtaining
\begin{figure}
\centering
\includegraphics[width=8cm]{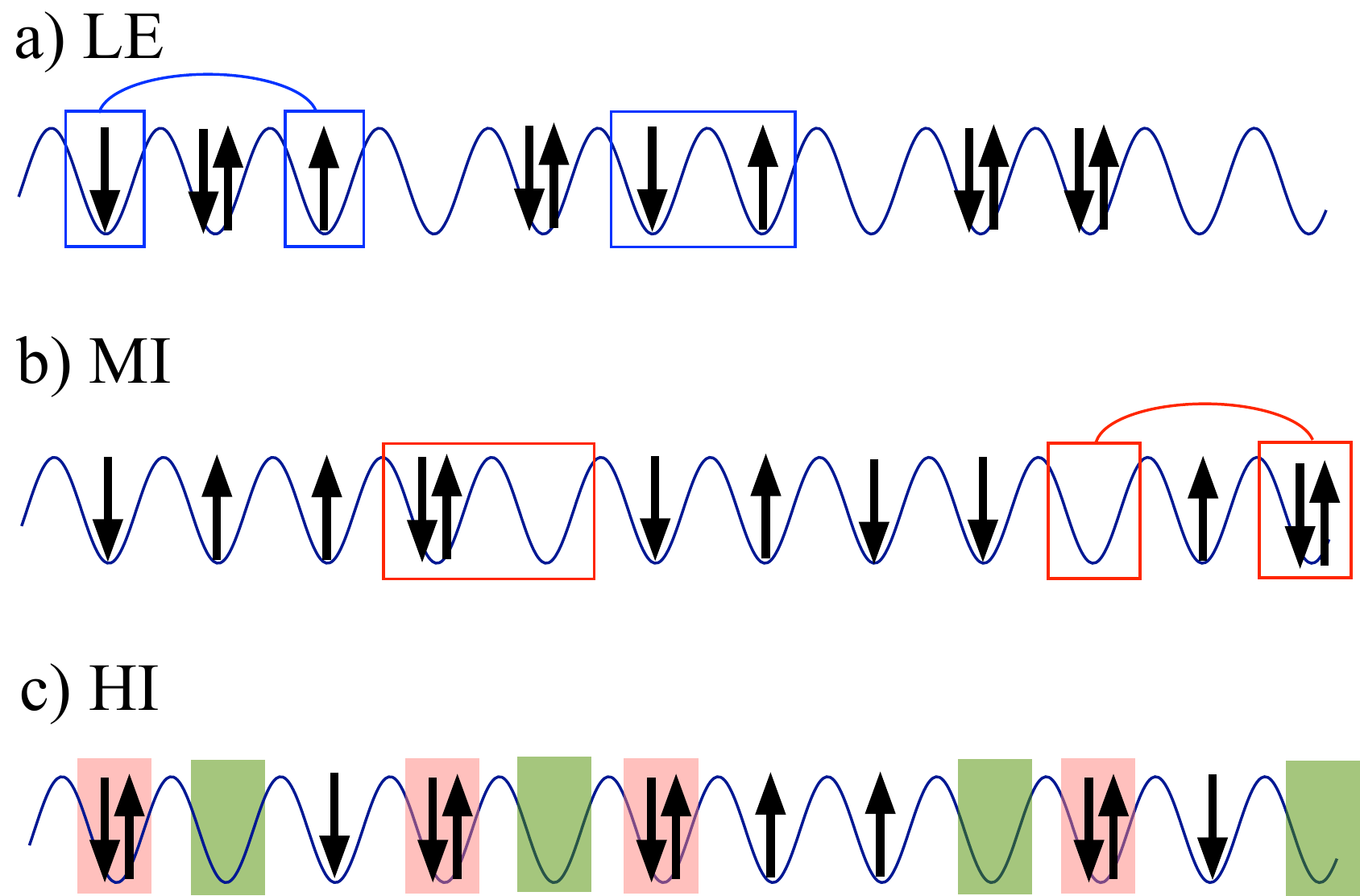}
\caption{(Color online)   Schematic representation  of various phases with a gap in either the spin or the charge sector, at half filling. a) the Luther Emery (LE) liquid phase exhibits a $C_P^{(s)}$ order, where pairs of singly occupied sites with spin-$\uparrow$ and spin-$\downarrow$ fermions are correlated and localized close to each other along the 1D lattice; b) the Mott-Insulator (MI) phase exhibits a $C_P^{(c)}$ order, where  correlated doubly occupied sites (doublons) and empty site (holons) are localized near to each other; c)   the Haldane-Insulator (HI) exhibits a non vanishing $C_S^{(c)}$ correlation, which implies that doublons and holons appear in alternate order in the sub-lattice of non-singly occupied sites. See Ref.[\onlinecite{BMR}] for details.}
\label{fig1}
\end{figure} 

\begin{align}
C_{P}^{(\nu)}(x) &= \langle \cos[\sqrt{2 \pi}\Phi_{\nu}(0)]\cos[\sqrt{2\pi }\Phi_{\nu}(x)] \rangle\label{eq:op}\\
C_{S}^{(\nu)}(x) &= \langle \sin[\sqrt{2 \pi}\Phi_{\nu}(0)]\sin[\sqrt{2\pi }\Phi_{\nu}(x)] \rangle \quad.\label{eq:os}
\end{align}
The different pinning values for the field $\Phi_\nu$ in table (\ref{table1})  determine the asymptotic behavior $\lim_{x \rightarrow \infty} C_{P/S}^{(\nu)} (x)$, leading to a  non-vanishing value of at least one of the correlations, determining the order parameters for the gapped phases. The  microscopic orders can be deduced from the analysis developed in Ref.[\onlinecite{BMR}] and are pictorially sketched   in  Fig.\ref{fig1} .  
Also, phases can  be further characterized by the asymptotic behavior of more customary local operators, which instead decay at least algebraically, such as
\begin{eqnarray}
O_{CDW}(x)=  \sin[\sqrt{2 \pi} \Phi_c(0)] \cos[\sqrt{2 \pi} \Phi_s(x)] \label{OCDW} \\
O_{SDW}(x)= \cos[\sqrt{2 \pi} \Phi_c(0)] \sin[\sqrt{2 \pi} \Phi_s(x)] \label{OSDW} \\
O_{TS}(x)=  \exp[i \sqrt{2 \pi} \Theta_c(0)] \sin[\sqrt{2 \pi} \Phi_s(x)] \label{OTS} \\
O_{SS}(x)= \exp[i \sqrt{2 \pi} \Theta_c(0)] \cos[\sqrt{2 \pi} \Phi_s(x)] \label{OSS}
\end{eqnarray}
where $SDW$ stands for spin density waves, $TS$ and $SS$ for triplet and singlet superconductivity respectively.

\subsection{Luttinger Liquid Phase}
In the Luttinger liquid (LL) phase both charge and spin sector are gapless. Evidence of LL behavior was found in both condensed matter systems~\cite{QW,CNT,QHE} and ultracold gases \cite{LLCA}. The correlation functions are characterized by quasi-long range order, i.e. they decay  with a power-law behavior  at large distance. The exponents of the power-laws are non-universal (in that they are  interaction-dependent), although the mutual relations between the exponents do determine a universality class.  \\
The LL phase  is present under the following conditions \\

\noindent i) {\it At half filling} ($\rho=1$)  the following two relations must be fulfilled\\
\begin{equation}   
\left\{ \begin{array}{l} 
\displaystyle U-2V-\frac{8 \tilde{X}}{\pi}-2Y+\frac{3}{2}J - \frac{4Q}{\pi^2} \le 0 \\
\displaystyle 8V+4Y+8P+Q(2+\frac{1}{\pi^2}) \le 0 \\
\displaystyle U-2V+ \frac{8 \tilde{X}}{\pi}+2Y-\frac{J}{2}+ \frac{4Q}{\pi^2} \ge 0
\end{array}
\right. \label{LL-i)}
\end{equation}
or 
\begin{equation}   
\left\{ \begin{array}{l} 
\displaystyle U-2V-\frac{8 \tilde{X}}{\pi}-2Y+\frac{3}{2}J - \frac{4Q}{\pi^2} \ge 0 \\
\displaystyle 2U+4V-\frac{16 \tilde{X}}{\pi} +3J +8P+Q(2-\frac{7}{\pi^2})\le 0\\
\displaystyle U-2V+ \frac{8 \tilde{X}}{\pi}+2Y-\frac{J}{2}+ \frac{4Q}{\pi^2} \ge 0
\end{array}
\right. \label{LL-ii)}
\end{equation}
In this case, the dominant correlation functions are the superconducting ones, and in particular the triplet is known to be logarithmically dominant with respect to the singlet\cite{VOIT,GIAM}; we thus have:
\begin{equation}
\langle O_{TS} (x) O^{\dagger}_{TS}(y) \rangle \sim |x-y|^{-(1+K_c^*)}
\end{equation}
where $K_c^*$ is the fixed-point value  
\begin{eqnarray}
K_c^{*}&=&1+  \frac{1}{2\pi t}  \left( \frac{}{} (4 V +2 Y + 4 P+Q(1+\frac{1}{2\pi^2})) \right.  \label{Kcstar} \\
& & \left. (U+2 V -\frac{8 \tilde{X}}{\pi}+\frac{3}{2} J + 4 P+Q(1- \frac{7}{\pi^2}) \right)^{1/2} \nonumber
\end{eqnarray}
For $V=\tilde{X}=J=P=Q=0$ one recovers the result of \cite{JAKA3} for the Penson-Kolb-Hubbard model. \\Notice that, differently from the ordinary Hubbard model at half-filling, the RG flux of the charge sector does not necessarily take place along a separatrix, because the extra interaction terms make the model not charge SU(2) symmetric. Indeed in general $K_c^* \neq 1$. \\

\noindent ii) {\it Away from half filling} ($\rho \ne 1$)  one is left with only one condition:
\begin{equation}
B_s(\rho) \, > \, 0
\end{equation}
where $B_s$ is given by the full expression (\ref{Bs}), generalizing 
  the result by Ref.[\onlinecite{AA}] to the case of non-vanishing $\tilde{X}, P$ and $Q$. In this case the dominant correlations function are still the TS ones. However
\begin{equation}
\langle O_{TS} (x) O^{\dagger}_{TS}(y) \rangle \sim |x-y|^{-(1+K_c)}
\end{equation}
where $K_c$ is the bare parameter given by (\ref{Knu}) and (\ref{Bc}).
\\\\
\subsection{Luther-Emery liquid phase}
The LE liquid phase is characterized by gapless charge excitations, and a gapped spin sector. This implies that the RG flow of the spin sector must take place along the outgoing separatrix, i.e. that $B_s=C_s <0$. Owing to that, the field $\Phi_s$ is pinned to one of the infinitely many degenerate minima of the potential $m_s \cos{\sqrt{8 \pi} \Phi_s}$ in Eq.(\ref{HBOS}), shown in table (\ref{table1}). Hence in the LE phase the LRO is described by the parity spin correlator $C_P^{(s)}$,  which remains finite in the thermodynamic limit. This phase is microscopically characterized by correlated pairs of singly occupied sites with spin-$\uparrow$ and spin-$\downarrow$ fermions that are localized, i.e. that are likely to be distributed in neighboring sites along the lattice [see Fig.\ref{fig1}a)]. 
\\The correlation functions of the local operators  (\ref{OCDW})-(\ref{OSS}) are instead difficult to evaluate in general, due to the gapped spin part. However, at the decoupling point $K_s=1/2$, they can be calculated exactly since the model can be refermionized into a free massive Dirac fermions\cite{LADDER}; we emphasize that, strictly speaking, such point is beyond the consistency condition of the weak-coupling approach, which implies that operators are marginal, i.e. that $K_s$ is always close to 1 [see Eqs.(\ref{Knu}) and (\ref{Bc})-(\ref{Bs})]. However, it is known from the exact solution\cite{BOOK,SGM} that the picture valid at $K_s=1/2$ is robust for the whole region $1/2 \le K_s <1$, and thus also for $K_s \lesssim 1$. In contrast, for $K_s < 1/2$  breathers (bound states) can appear, and the form factor  approach has to be invoked\cite{SGMff}.  The SDW and TS correlation functions decay exponentially fast, whereas the CDW and the SS exhibit a power-law behavior (due to the charge sector) whose exponent depend on whether the system is half-filled or not. The phase exists under the following conditions:\\

\noindent {\sl At half filling} ($\rho=1$)  the following two relations must be fulfilled\\
\begin{equation}   
\left\{ \begin{array}{l} 
\displaystyle U-2V-\frac{8 \tilde{X}}{\pi}-2Y+\frac{3}{2}J - \frac{4Q}{\pi^2} \le 0 \\
\displaystyle 8V+4Y+8P+Q(2+\frac{1}{\pi^2}) \le 0 \\
\displaystyle U-2V+ \frac{8 \tilde{X}}{\pi}+2Y-\frac{J}{2}+ \frac{4Q}{\pi^2} < 0
\end{array}
\right. \label{LL-i)}
\end{equation}
or 
\begin{equation}   
\left\{ \begin{array}{l} 
\displaystyle U-2V-\frac{8 \tilde{X}}{\pi}-2Y+\frac{3}{2}J - \frac{4Q}{\pi^2} \ge 0 \\
\displaystyle 2U+4V-\frac{16 \tilde{X}}{\pi} +3J +8P+Q(2-\frac{7}{\pi^2})\le 0\\
\displaystyle U-2V+ \frac{8 \tilde{X}}{\pi}+2Y-\frac{J}{2}+ \frac{4Q}{\pi^2} < 0
\end{array}
\right. \label{LL-ii)}
\end{equation}
and the dominant order parameters are
\begin{equation}
\langle O_{CDW} (x) O^{\dagger}_{CDW}(y) \rangle = \langle O_{SS} (x) O^{\dagger}_{SS}(y) \rangle \sim |x-y|^{-1/{K_c^*}}
\end{equation}
where $K_c^*$ is given by (\ref{Kcstar}).\\

\noindent {\sl Away from half filling} ($\rho \ne 1$)  one is left with only one condition:
\begin{equation}
B_s(\rho) \, < \, 0
\end{equation}
In this case one obtains
\begin{equation}
\langle O_{CDW} (x) O^{\dagger}_{CDW}(y) \rangle = \langle O_{SS} (x) O^{\dagger}_{SS}(y) \rangle \sim |x-y|^{-1}
\end{equation}
\subsection{Charge Insulator Phases}
When the charge sector is gapped and the  spin sector flows to the gapless fixed point $K_s^*=1$, the system behaves as a charge insulator.
Such situation occurs only at  half filling. 
In this case the charge field $\Phi_c$ is pinned. For $m_c < 0$, one has $\Phi_c=p_c \pi$ ($p_c \in \mathbb{Z}$) as  pinning values, and $C_P^{(c)}$ remains finite. This is the MI   phase,  which is characterized by correlated pairs of doublons and holons,  localized near to each other [see Fig.\ref{fig1}b)], and where  SDW correlations are dominant  
\begin{equation}
\langle O_{SDW} (x) O^{\dagger}_{SDW}(y) \rangle \sim |x-y|^{-1} \quad.
\end{equation}
In contrast, for $m_c > 0$, the pinning value is $\sqrt{2\pi} \Phi_c   = \pi (p_c+1/2)$ ($p_c \in \mathbb{Z}$). This is the HI   phase, where LRO is described by the finite value of $C_S^{(c)}$, and  CDW correlations turn out to be dominant
\begin{equation}
\langle O_{CDW} (x) O^{\dagger}_{CDW}(y) \rangle \sim |x-y|^{-1}
\end{equation}
The microscopic order  amounts to correlated doublons and holons, which appear in alternated order [see Fig.\ref{fig1}c)]. In this case  CDW correlations are dominant. \\

Explicitly, the MI phase occurs for
\begin{equation}   
\left\{ \begin{array}{l} 
\displaystyle U-2V-\frac{8 \tilde{X}}{\pi}-2Y+\frac{3}{2}J - \frac{4Q}{\pi^2} > 0 \\
\displaystyle 2U+4V-\frac{16 \tilde{X}}{\pi} +3J +8P+Q(2-\frac{7}{\pi^2})>0\\
\displaystyle U-2V+ \frac{8 \tilde{X}}{\pi}+2Y-\frac{J}{2}+ \frac{4Q}{\pi^2} \ge 0
\end{array}
\right. \label{CI-i)}
\end{equation}
whereas the HI phase is realized for
\begin{equation}   
\left\{ \begin{array}{l} 
\displaystyle U-2V-\frac{8 \tilde{X}}{\pi}-2Y+\frac{3}{2}J - \frac{4Q}{\pi^2} < 0 \\
\displaystyle 8V+4Y+8P+Q(2+\frac{1}{\pi^2}) >0 \\
\displaystyle U-2V+ \frac{8 \tilde{X}}{\pi}+2Y-\frac{J}{2}+ \frac{4Q}{\pi^2} \ge 0
\end{array}
\right. \label{CI-ii)}
\end{equation}

\subsection{Fully gapped phases}
This type of phases, which can occur only at half filling $\rho=1$, is characterized by both massive channels, so that both fields are pinned. In particular, since $m_s<0$, the field $\Phi_s$ is always pinned around the values $ \sqrt{2 \pi} \Phi_s   = p_s \pi$ ($p_s \in \mathbb{Z}$),  so that $C_P^{(s)}$ is finite. Moreover, depending on the sign of $m_c$, two possible sets of pinning values are possible for $\Phi_c$,  giving rise to  two different types of LRO  in the charge sector.  
When $m_c<0$, the field $\Phi_c$ is pinned around $\sqrt{2 \pi} \Phi_c   =p_c \pi$ ($p_c \in \mathbb{Z}$). In this case $C_P^{(c)}$ is also finite, and the microscopic order consists of correlated pairs of  doublons and holons and correlated pairs of singly occupied sites with spin-$\uparrow$ and spin-$\downarrow$ fermions, that are likely to be distributed in neighboring sites. The dominant correlations are of BOW type. In contrast,  if $m_c>0$ the pinning value is $\sqrt{2 \pi}   \Phi_c   =\pi (p_c+1/2)$, and the phase is characterized by a finite $C_S^{(c)}$, besides a finite $C_P^{(s)}$. The microscopic order thus amounts to correlated pairs of singly occupied spin-$\uparrow$ and spin-$\downarrow$ sites, that are localized near each other in a background of  alternated doublons and holons; CDW correlations are dominant.
Notice that, in both phases, singly occupied sites are localized close to each other, whereas holons and doublons can either appear in localized pairs (BOW), or in alternate order along the chain  (CDW). The presence of LRO in the fully gapped phases is usually envisaged through the finite asymptotic value of the CDW and BOW correlation functions.  Indeed the analysis at the decoupling points $K_c=K_s=1/2$ shows that
\begin{equation}
\langle O_{CDW} (x) O^{\dagger}_{CDW}(y) \rangle  \sim  \mbox{const} \quad.
\end{equation}
Already in case of the half-filled $U-V$ model with $U<2V$ the above requirements are fulfilled; such a long-range order is related to the breaking of a {\it discrete} symmetry (the translation by one site) in the insulating ground state. On the other hand, in a typical compound one can have at most $U \lesssim 2V$; therefore, although $U$ and $V$ are the most relevant coupling constants, the other interaction terms such as $\tilde{X}, Y, J, P$ and $Q$ can occur to be of the order of $U-2V$; the present results quantitatively point out that the latter can determine the presence  of the above long-range order.\\\\
Here below we provide the conditions at half-filling for arbitrary parameter values.
A fully gapped CDW occurs for

\begin{equation}   
\left\{ \begin{array}{l} 
\displaystyle U-2V-\frac{8 \tilde{X}}{\pi}-2Y+\frac{3}{2}J - \frac{4Q}{\pi^2} < 0 \\
\displaystyle 8V+4Y+8P+Q(2+\frac{1}{\pi^2}) >0 \\
\displaystyle U-2V+ \frac{8 \tilde{X}}{\pi}+2Y-\frac{J}{2}+ \frac{4Q}{\pi^2} <0 
\end{array}
\right. \label{FG-i)}
\end{equation}
whereas a fully gapped BOW phase occurs for
\begin{equation}   
\left\{ \begin{array}{l} 
\displaystyle U-2V-\frac{8 \tilde{X}}{\pi}-2Y+\frac{3}{2}J - \frac{4Q}{\pi^2} > 0 \\
\displaystyle 2U+4V-\frac{16 \tilde{X}}{\pi} +3J +8P+Q(2-\frac{7}{\pi^2})>0\\
\displaystyle U-2V+ \frac{8 \tilde{X}}{\pi}+2Y-\frac{J}{2}+ \frac{4Q}{\pi^2} < 0
\end{array}
\right. \label{FG-ii)}
\end{equation}

\section{Effects of diagonal three- and four-body interactions}
\label{sec-4}
The experimental realization of confinement of ultracold gases of multiple species and non-vanishing dipolar moment has opened the way to the engineering of many body interactions of order higher than two\cite{BMZ,Bonnes,hammer}. Signature of three- and four-body interactions have recently been experimentally observed in systems of Rb and Cs atoms in an optical lattices~\cite{han,gurian}.  In bosonic systems, three-body  terms  have been shown~\cite{BMZ,CSal} to lead to a super-solid phase,  characterized by the simultaneous presence of charge modulations and superconducting correlations,   at appropriate commensurate fillings. For  fermionic systems, the three-body couplings that have been mostly analyzed are correlated hopping terms, characterized by the coupling constant   $\tilde{X}$ in Eq.(\ref{EHM}). These terms were first considered in the field of superconductivity of narrow-band materials~\cite{HIRSCH,AAch,AAS,AA}, and have more   recently been investigated in the context of cold atoms. In particular, it has recently been predicted that the three-body coupling $\tilde{X}$ can be responsible for the appearance of Haldane charge order at half-filling~\cite {BMR}. Such type of three-body couplings \cite{nakamura} are off-diagonal in the occupation number representation. However,
 most of setups of ultra cold gases involve diagonal many-body terms~\cite{BMZ,DIDE}, i.e. terms  that directly couple the local electron density $\hat{n}_{j\sigma}$ at each lattice site. In the lattice Hamiltonian (\ref{EHM}) such diagonal three-body  and four-body  terms are characterized by the coupling constants $P$  and $Q$, respectively, and represent the natural generalization of the conventional diagonal two-body couplings $U$ and $V$. We shall thus now specify the general results obtained in previous sections to analyze the phase diagram in the case where only $U$, $V$, $P$ and $Q$ couplings are present. In particular, because the analysis as a function of the  coupling $V$ has already been widely explored in the literature\cite{voitPRB,kolomeisky,nakamura}, we shall address here the effects of the three and four body coupling $P$ and $Q$. \\
\begin{figure}
\centering
\includegraphics[width=7cm]{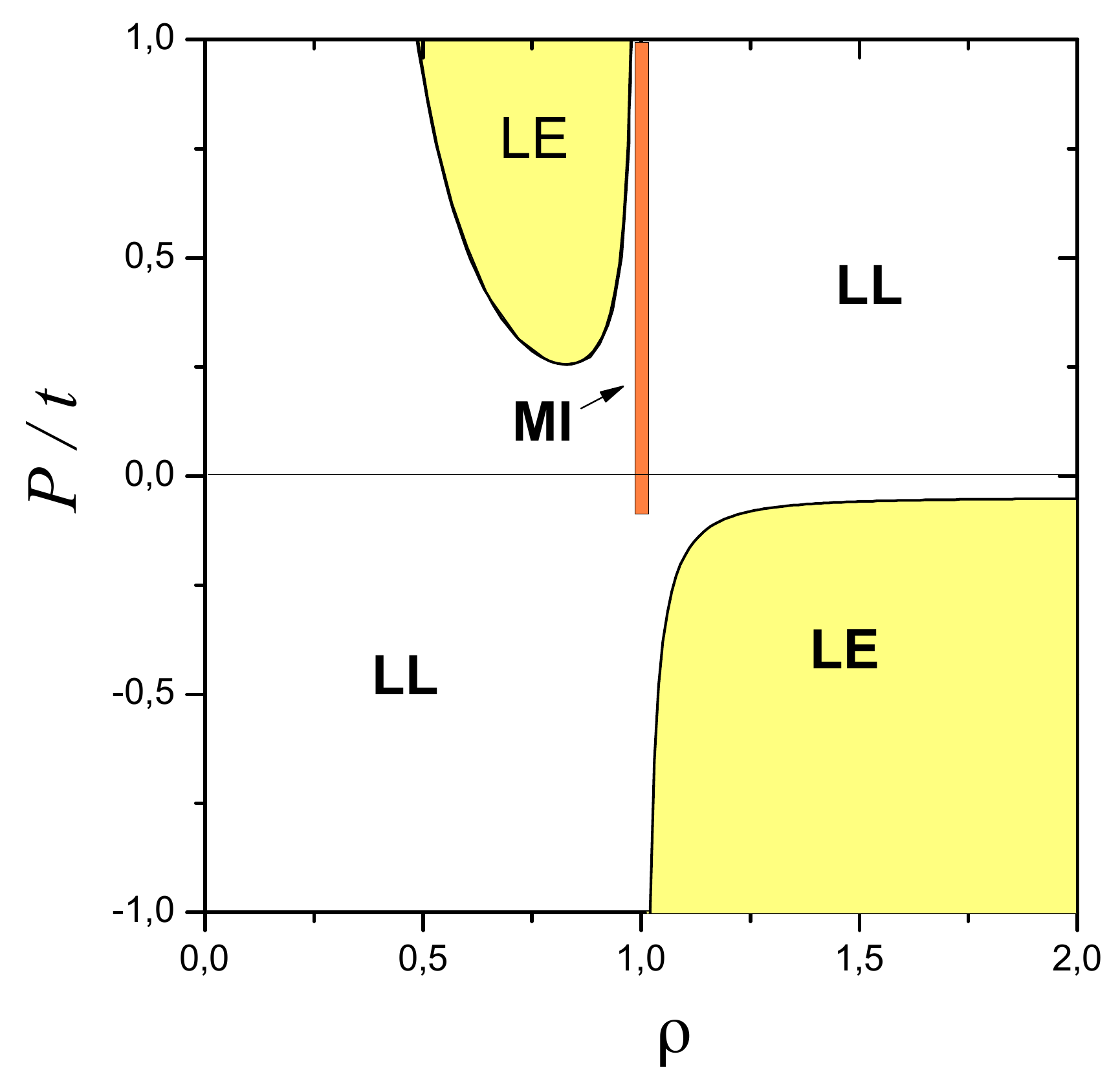}
\caption{(Color online) The phase diagram of the Extended Hubbard model for $U=3V=t/4$ as a function of the filling factor $\rho$ and the three-body diagonal coupling $P$. Transition from LL to LE phases are possible both for positive and negative values of $P$, depending on the filling factor. At half filling $\rho=1$, the   MI transition occurs for both repulsive and moderately attractive three-body coupling $P$.}
\label{phase-diagram_rho-P}
\end{figure} 

\begin{figure}
\centering
\includegraphics[width=7cm]{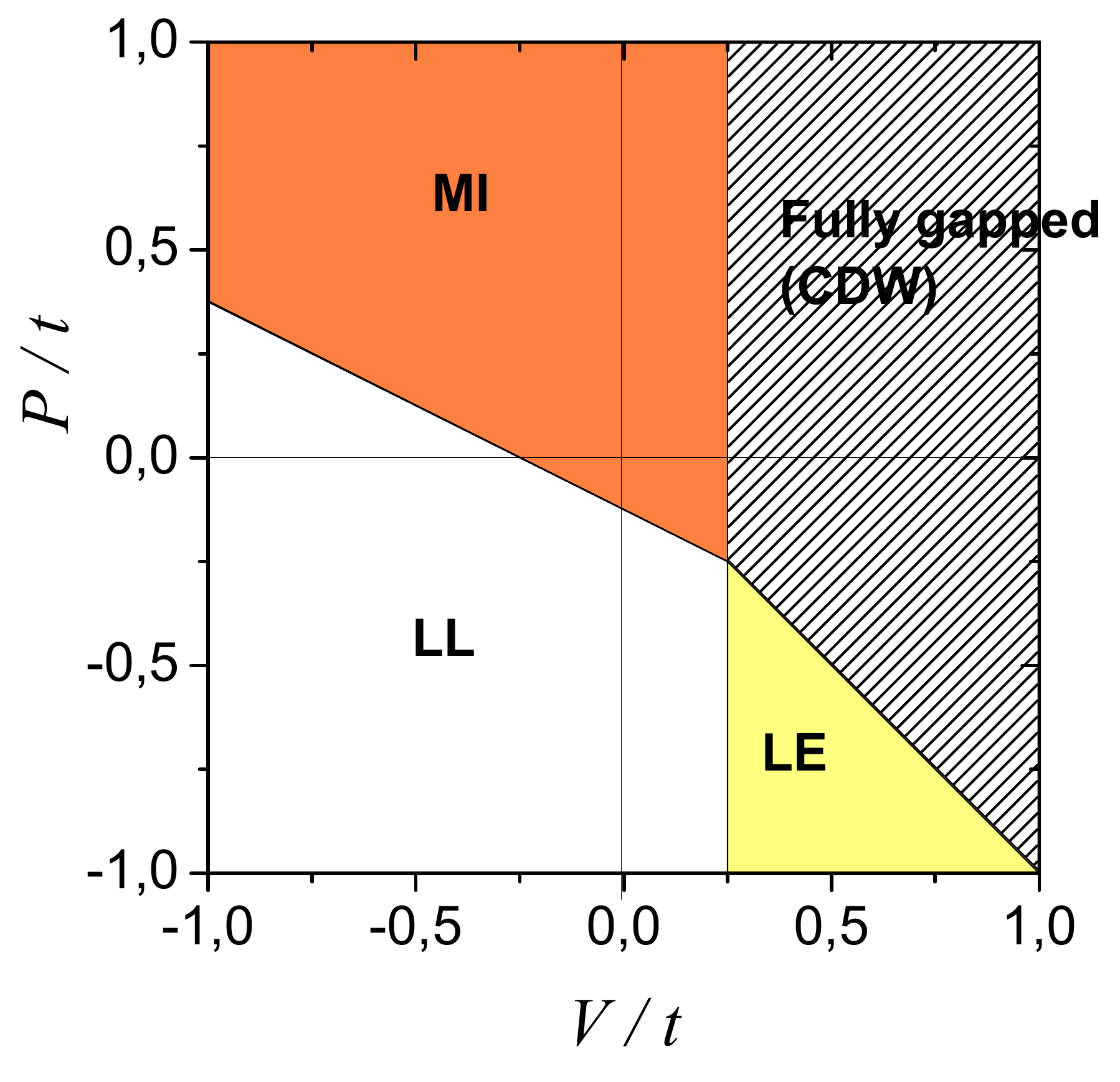}
\caption{(Color online) Ground state phase diagram of the extended Hubbard model (\ref{EHM}) as a function of the two-body nearest neighbors interaction parameter $V$ and of the diagonal three-body coupling $P$, for $U=t/2$ at half-filling $\rho=1$. For $V<U/2$, the  $P$   coupling  drives a transition from a LL to MI phase, whereas for  $V>U/2$  the $P$ coupling drives a transition from a fully gapped CDW phase to a LE phase, characterized by a gapless charge sector and a gapped spin sector. Differently from the $U-V$ extended Hubbard model, the $P$ term makes the LE phase arise also for repulsive $U,V>0$. This phase diagram has to be compared with the one of the off-diagonal  three-body  coupling $\tilde{X}$ originating from correlated hopping in Eq.(\ref{EHM}), shown in Fig.\ref{Fig_VXtilde}.}
\label{Fig_VP}
\end{figure}
\begin{figure}
\centering
\includegraphics[width=7cm]{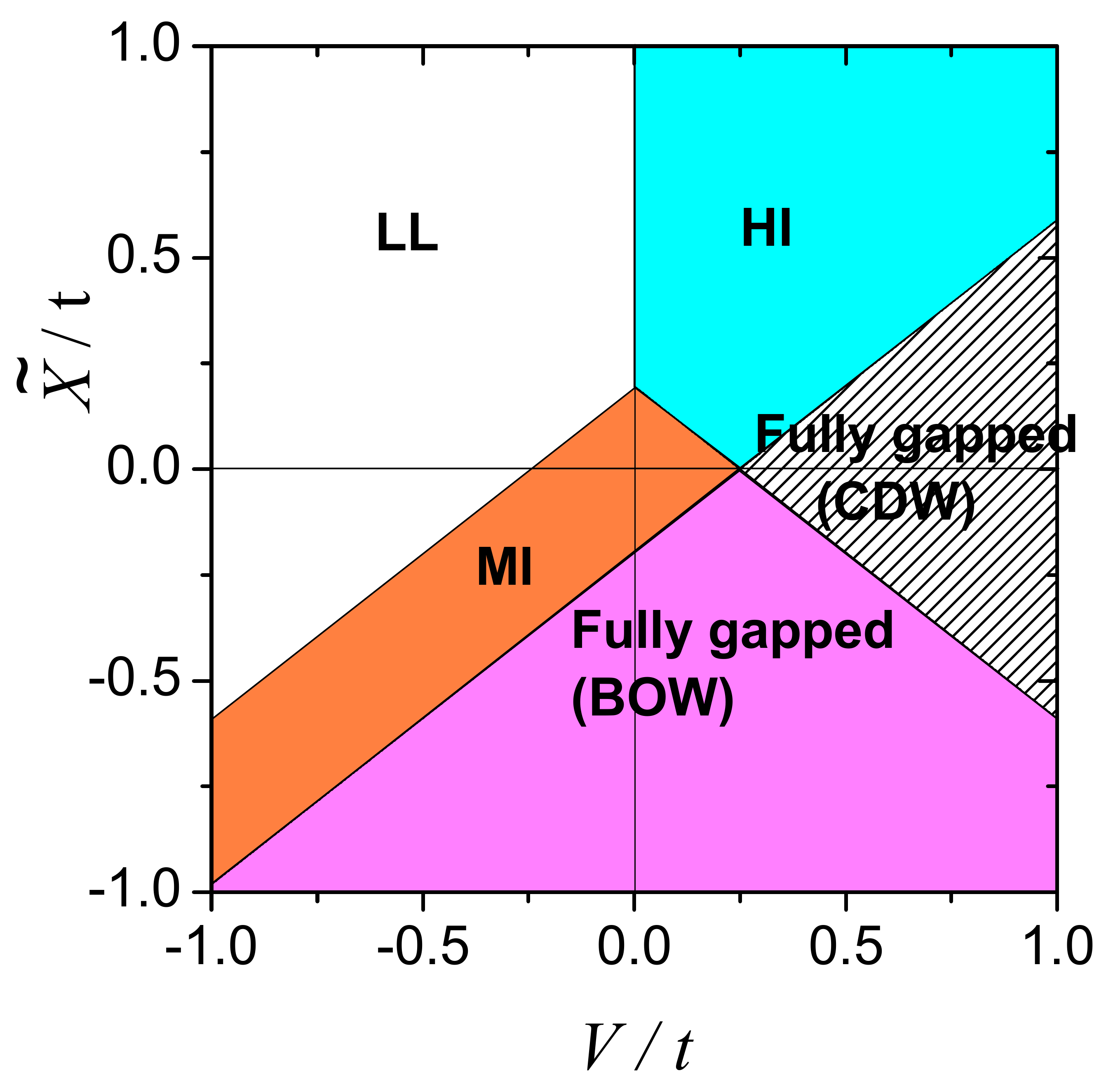}
\caption{(Color online) Ground state phase diagram of the extended Hubbard model (\ref{EHM}) as a function of the two-body nearest neighbors interaction parameter $V$ and of the off-diagonal three-body coupling $\tilde{X}$, for $U=t/2$ at half-filling $\rho=1$. Differently from Fig.\ref{Fig_VP}, no LE phase is present. Instead, when $V>0$, a HI phase, characterized by a gapless spin sector and a gapped charge sector, emerges. Furthermore, for $V>U/2$, the fully gapped CDW phase, already present in the $U-V$ model, persists. }
\label{Fig_VXtilde}
\end{figure}

\subsection{Three-body interaction}
We start from analyzing the effect of the three-body coupling $P$, and set first $Q=0$. In Fig.\ref{phase-diagram_rho-P}  the ground state phase diagram of the Extended Hubbard model with $U=3V=t$ is plotted as a function of the three-body term $P$ and the filling factor $\rho$. As one can see, LE phases appear for both repulsive and attractive values of $P$, namely for $P>0$ at $\rho<1$ and for $P<0$ at $\rho>1$. With varying the filling factor, transition from LL to a LE phase occur for both positive and negative $P$ values. 
At half filling  ($\rho=1$) the $P$ term changes the threshold values of $U$ and $V$  for the onset of MI phase, which appears  when $U+2V>-4P$ and $U-2V>0$. Notice that, for suitable values of the two-body couplings $U$ and $V$, a repulsive diagonal three-body term $P>0$ makes the MI phase in principle possible even when $U$ and $V$ are  both attractive ($U,V<0$). \\
The case of half filling is particularly suitable to highlight the different roles played by the diagonal three-body coupling $P$ and the off-diagonal three-body coupling $\tilde X$, originating from correlated hopping [see Eq.(\ref{EHM})] previously considered in the literature (see e.g. Ref.[\onlinecite{AA,nakamura}]).  In particular Fig.~\ref{Fig_VP} shows the phase diagram as a function of $V$ and $P$, for repulsive on-site coupling $U=t/2>0$. As one can see,  for $V<U/2$ the three-body term $P$ induces a transition between the LL phase and a MI phase, whereas for $V>U/2$ the coupling $P$ drives a transition from the LE phase into the fully gapped phase with CDW order.  An attractive value $P<0$ of the three-body coupling makes the LE phase appear even for repulsive  two-body couplings, $U,V>0$. Importantly, this effect is absent in the $U-V$ extended Hubbard model with $U>0$, regardless of the sign of $V$, and cannot be induced by the off-diagonal three-body coupling $\tilde X$ either. This is illustrated in Fig.\ref{Fig_VXtilde}, where the phase diagram is plotted as a function of $V$ and $\tilde{X}$, for the same repulsive on-site coupling $U=t/2>0$ as Fig.\ref{Fig_VP}. At repulsive $V>0$, depending on the sign of $\tilde{X}$, a fully gapped BOW phase or a charge insulator HI phase emerges. The latter was previously known in the literature as BSDW~\cite{nakamura}. Furthermore, for $V>U/2$, the fully gapped CDW phase, already present in the $U-V$ model, persists. \\The direct inspection of Figs.~\ref{Fig_VP} and \ref{Fig_VXtilde} emphasizes that, while the off-diagonal three body coupling $\tilde{X}$ favors the emergence of the HI phase (charge sector gapped, spin sector gapless), the diagonal three-body coupling $P$ favors the LE phase (charge sector gapless, spin sector gapped). We also notice that, while in Fig.~\ref{Fig_VP} all transitions are of BKT type, in the case of Fig.\ref{Fig_VXtilde} a second order transition line $8\tilde{X}/\pi= U-2 V$ emerges for $V>0$, separating MI and HI, and BOW and CDW phases.
 
\subsection{Four-body interactions}
\begin{figure}
\centering
\includegraphics[width=8cm]{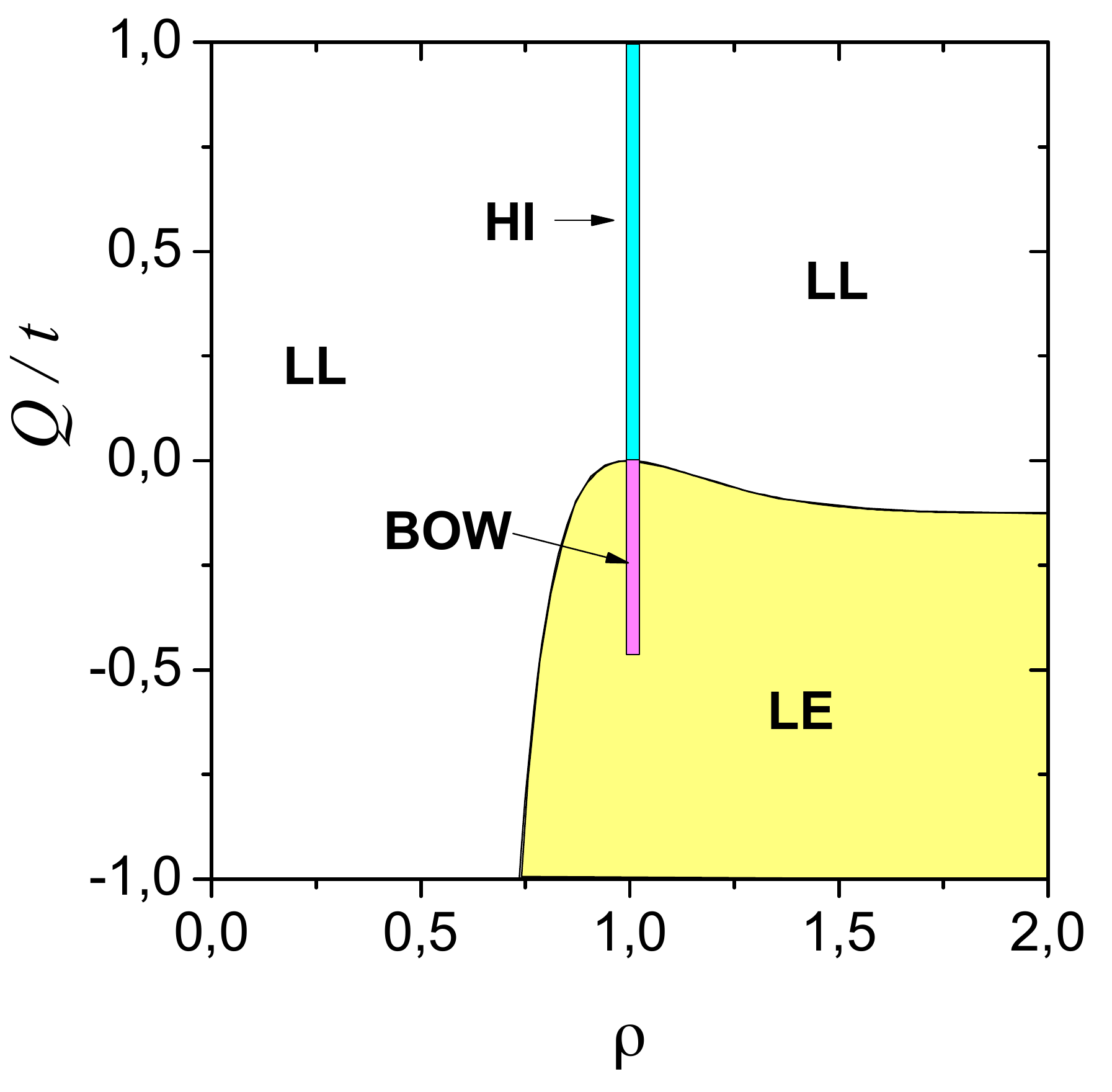}
\caption{(Color online) The phase diagram of the Extended Hubbard model for $U=2V=t/4$ as a function of the filling factor $\rho$ and the four-body diagonal coupling $Q$. Transition from LL to LE phases are possible both for positive and negative values of $Q$, depending on the filling factor. At half filling $\rho=1$, two transitions occur  from a LE to a fully gapped BOW phase, and from the BOW phase to a   MI phase, with varying the four-body coupling $Q$ from attractive to repulsive regime.}
\label{phase-diagram_rho-Q}
\end{figure}

\begin{figure}
\centering
\includegraphics[width=8cm]{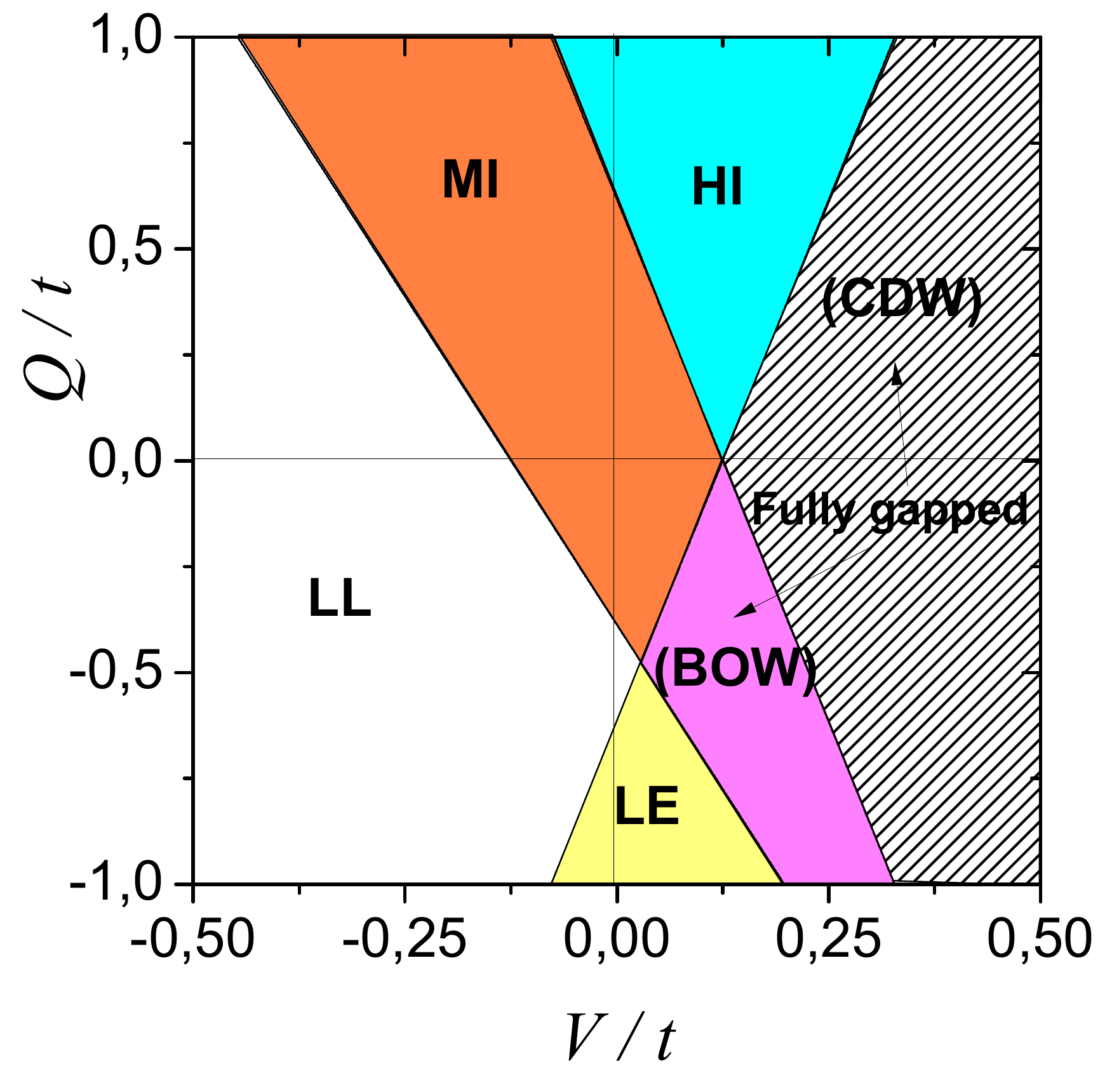}
\caption{(Color online) Ground state phase diagram of the model in the $V-Q$ space for $U=t/4$, at half-filling $\rho=1$. All possible phases described in table (\ref{table1}) appear when varying the two-body coupling $V$ and the four-body coupling $Q$. In particular, one observes the presence of a Haldane Insulator phase for repulsive $U,V,Q>0$.}
\label{phase-diagram_V-Q}
\end{figure}

Let us now consider the effect of the four-body coupling $Q$ appearing in Eq.(\ref{EHM}), and set $P=0$. The phase diagram as a function of the filling  $\rho$ and $Q$, for the case $U=2V=t$, is shown in Fig.\ref{phase-diagram_rho-Q}. As one can see, at $Q<0$ a filling-driven transition between a LL and a LE phase occurs. In particular, at half-filling,  transitions from LE to a fully gapped BOW phase and to a HI phase occur with varying the coupling $Q$.
In fact, at half-filling the situation turns out to be particularly interesting because of the possible opening of the charge gap. In Fig.\ref{phase-diagram_V-Q}  the phase diagram is plotted as a function of the two-body coupling $V$ and the four-body coupling $Q$,  for $U=t/4$, and exhibits an extremely rich structure, where all possible phases identified in Table (\ref{table1}) can be observed already at half filling. This confirms at a glance the interesting role played by such diagonal-four body interaction. \\
Let us in particular discuss the  charge insulator phases, HI and MI, whose parameter conditions are detemined by Eqs. (\ref{CI-i)}), (\ref{CI-ii)}). One can see that the presence of Haldane order in the charge sector is favored by a repulsive four-body term $Q>0$, while a repulsive two-body term $U>0$  favors Mott (parity) order.  Indeed the two different charge orders are induced by different arrangement of doublons and holons in the background of singly occupied sites (see Fig.\ref{fig1}). The result corresponds to the physical intuition that a repulsive~$Q$  prevents the formation of neighboring pairs of doublons,  a  feature that is favored by the alternation of doublons and holons characterizing Haldane order. The direct observation of Haldane order in low dimensional fermionic systems has remained an open issue so far, since previous theoretical investigations have suggested that off-diagonal terms are necessary to observe it \cite{BMR,nakamura}. However, this type of coupling is difficult to realize experimentally\cite{LEW}. Our result suggests that the observation of Haldane charge order in trapped ultra cold gases of fermionic atoms is possible, upon inducing a diagonal four-body interaction term. Also, a second order transition line is observed between the HI and MI phases, as well as between BOW and CDW phases. Even at $V=0$ the independent tuning of $U$ would allow the observation of the direct second order MI to HI transition at $U=4 Q/\pi^2$.  Such features should be present also in bosonic case, since they do not appear to be related to the presence of spin degree of freedom.

\section{Conclusions}
\label{sec-5}
We  have applied the Bosonization technique to investigate a widely general class of extended Hubbard models [see Eq.(\ref{EHM})], which includes a variety of two-body couplings up to nearest neighboring sites, and also three- and four-body interaction terms. These models, which now find a promising platform in gases of ultracold dipolar molecules trapped in optical lattices, also describe several  physical features  of  1D materials in condensed matter. We have determined the relations that coupling constants appearing in Eq.(\ref{EHM}) must fulfill for the opening/closing of the charge and spin gap, thereby characterizing the conditions for the emergence of LL, LE, CDW, BOW, HI and MI phases. \\We have then focussed our investigation on the effects of diagonal three- and four-body couplings, characterized by the coupling constants $P$ and $Q$ in Eq.(\ref{EHM}). We have proved that these terms, whose realization in systems of interacting dipolar molecules \cite{BMZ} is  nowadays at experimental reach,  have   non-trivial effects  on the phase diagram of the system. Our analysis has been carried out at {\it arbitrary} filling $\rho$, and has determined  the existance of filling dependent phase boundaries between LL and LE phases, as shown in Figs.~\ref{phase-diagram_rho-P} and \ref{phase-diagram_rho-Q}. \\A quite appealing scenario occurs at half-filling ($\rho=1$), where a gap may open in the charge sector, depending on the values of the various coupling constants.  
In particular, we have found that the three-body term $P$ induces a transition between the LL phase and a MI phase if $V<U/2$, whereas it determines a transition from the LE phase to the CDW phase for $V>U/2$.  Interestingly, an attractive value $P<0$ of the three-body coupling makes the LE phase appear even for repulsive  two-body couplings, $U,V>0$. Importantly, this effect is absent in the $U-V$ extended Hubbard model with $U>0$  and cannot be induced by the off-diagonal three-body coupling $\tilde X$, originating from the correlated hopping term that was previously investigated. Indeed  our results (see Figs.~\ref{Fig_VP} and \ref{Fig_VXtilde}) show  that, while the off-diagonal three body coupling $\tilde{X}$ favors the emergence of the HI phase (charge sector gapped, spin sector gapless), the diagonal three-body coupling $P$ favors the LE phase (charge sector gapless, spin sector gapped). Typically, off-diagonal couplings are more difficult to implement experimentally as compared to diagonal terms. This would suggest  that HI phase is unlikely to be observed. However, a possible way out to observe HI phase is offered by  the four-body coupling $Q$, which turns out to play  an extremely interesting role. Indeed, our result show that such term, in combination with the two-body density-density coupling $V$,   induces a quite rich phase diagram (see Fig.\ref{phase-diagram_V-Q})    where all possible phases can be present. In particular, also a HI phase is present for repulsive $U,V,Q>0$. Moreover a second order transition line emerges (separating HI from MI and CDW from BOW phases in Fig.\ref{phase-diagram_V-Q}). This is thus a different feature with respect to the case of three-body interactions, where such line occurs only in the presence of off-diagonal couplings.\\

A natural development of the present work would be to relax the constraint of SU(2) symmetry characterizing the spin-sector, by including a spin-orbit coupling~\cite{fuji-kawa}, whose effects on the ordinary Hubbard model have recently attracted a remarkable interest~\cite{SO-HUB}, especially in view of the realization of topological states using cold atoms systems~\cite{SO-LEW}. We expect that the interplay between spin-orbit coupling and three- and four-body terms might give rise to exotic phases, due to the much richer physics related to the spin sector. \\ 
As a final remark, we also mention that recent studies have pointed out that, when the weak coupling limit is abandoned, some qualitatively different results may be obtained. It was noticed~\cite{ADMO, AAA}, for instance, that in the half-filled bond-charge Hubbard model (where only  $U$ and $X$ terms are non-vanishing)  at moderate positive $U$   a transition from a MI to a BOW and then to a LE phase si driven by a sufficiently large $X$ term. This effect is not captured by the present low energy analysis, which predicts no effect of $X$ at $\rho=1$ in Eq. (\ref{Bs}). The result can be recovered within the   bosonization scheme by including higher order terms with respect to standard treatment (see also Ref.[\onlinecite{AD}]).  Hence,   another possible evolution of the present work may be the inclusion of such higher order terms in the Bosonization  approach. Also, accounting for Umklapp processes for multi particle scattering would allow the investigation of the conditions for charge gap opening also at commensurate fillings different from $\rho=1$, with the possible formation of Haldane and super solid phases~\cite{BSRG}. \\

\acknowledgments

We acknowledge interesting discussions with L. Arrachea, L. Barbiero, and M. Roncaglia; F.D. is particularly thankful to Prof. A. Nersesyan for illuminating discussions and suggestions, and  acknowledges financial support from FIRB 2012 project ÓHybridNanoDevÓ (Grant No.RBFR1236VV).

\appendix

\section{Low energy Hamiltonian}

\label{App}
Here we would like to provide some technical details concerning the procedure to obtain the low energy Hamiltonian (\ref{Hcs})-(\ref{HBOS}) from the original Hamiltonian (\ref{EHM}). In the first instance,  before performing the continuum limit (\ref{climit}), we have singled out  fluctuations $: \hat{n}_{j \sigma} :$ from the Fermi sea, by rewriting each density operators $\hat{n}_{j \sigma}$ appearing in~(\ref{EHM}) as  $\hat{n}_{j \sigma}= : \hat{n}_{j \sigma} : + \rho /2$ (where $\rho$ is the electron filling). This   avoids unphysical divergencies arising from the continuum limit of the density, and enables to bosonize $:\hat{n}_{j \sigma}:$ rigorously. In addition, the Operator Product Expansion (OPE)   has been applied in order to evaluate the fusion of fields in nearest neighboring sites, and in particular the following OPE formulas have been used
\begin{eqnarray}
\frac{
e^{\pm i \sqrt{4 \pi} \Phi_{\sigma}(x)}  \,
e^{\mp i \sqrt{4 \pi} \Phi_{\sigma}(x+a)}}
{(2 \pi \alpha)^2} & \simeq &
\frac{1}{(2 \pi a)^2} \mp i \frac{\partial_x \Phi_\sigma}{2 \pi^{3/2} a} - \\
&-&\frac{:(\partial_x \Phi_\sigma)^2 :}{2 \pi} \mp
\frac{i  \partial_x^2 \Phi_\sigma}{4 \pi^{3/2}} \nonumber \\
\frac{\partial_x \Phi_\sigma (x)}{\sqrt{\pi}}
\frac{e^{\pm i \sqrt{4 \pi} \Phi_\sigma(x+a)}}{2 \pi \alpha} &\simeq & \pm
\frac{i \,  e^{\pm i \sqrt{4 \pi} \Phi_\sigma(x)}}{2 \pi^2 a \,  \alpha}- \\
&-& \frac{: e^{\pm i \sqrt{4 \pi} \Phi_\sigma(x)} \partial_x \Phi_\sigma(x) :}
{L \sqrt{\pi}} \nonumber \\
\frac{e^{\pm i \sqrt{4 \pi} \Phi_\sigma(x)}}{2 \pi \alpha} \frac{\partial_x \Phi_\sigma (x+a)}{\sqrt{\pi}}   &\simeq & \mp
\frac{i \,  e^{\pm i \sqrt{4 \pi} \Phi_\sigma(x)}}{2 \pi^2 a \,  \alpha}+ \\
&+& \frac{: e^{\pm i \sqrt{4 \pi} \Phi_\sigma(x)} \partial_x \Phi_\sigma(x) :}
{L \sqrt{\pi}} \nonumber
\end{eqnarray}
where $L=N_s a$ is the length of the chain, and $: \, \, :$ stands henceforth for (bosonic) normal ordering. The field $\Phi_{\sigma}$ is chosen to fulfill periodic boundary conditions $\Phi_{\sigma}(L)=\Phi_{\sigma}(0)$, and we have considered $L\rightarrow \infty$.


\begin{references}
\bibitem{NOTE} In the notation of Ref.[\onlinecite{SAG}], we have $P=2(V_{12}-V_{11})$ and $Q=4(V_{11}+V_{22}-2 V_{12})$.
\bibitem{nota-RG} An inspection of Eqs.(\ref{RGeq}) shows that, in fact, the condition $\eta_\nu=0$ and $\xi_\nu<0$ also corresponds to   gapless fixed points. However, it is a half-line of   unstable fixed points, and any infinitesimal $\eta_\nu\neq 0$ leads the system to a gapped phase.
\end{references}
\end{document}